\documentstyle[preprint,aps]{revtex}

\begin{document}

\draft

\preprint{November,2000}

\title{Instanton-Meron Hybrid in the Background of \\
       Gravitational Instantons}

\author{Hongsu Kim\footnote{e-mail : hongsu@hepth.hanyang.ac.kr} and
Yongsung Yoon\footnote{e-mail : cem@hepth.hanyang.ac.kr}}

\address{Department of Physics \\
Hanyang University, Seoul, 133-791, KOREA}


\maketitle

\begin{abstract}
When it comes to the topological aspect, gravity may have profound effects
even at the level of particle physics despite its negligibly small relative
strength well below the Planck scale. In spite of this intriguing possibility,
relatively little attempt has been made toward the exhibition of this phenomenon
in relevant physical systems. In the present work, perhaps the simplest and 
the most straightforward new algorithm for generating solutions to (anti) 
self-dual Yang-Mills (YM) equation in the typical gravitational instanton 
backgrounds is proposed and then applied to find the solutions practically
in all the gravitational instantons known. Solutions thus obtained turn out to 
be some kind of instanton-meron hybrids possessing mixed features of both.
Namely, they are rather exotic type of configurations obeyng first order
(anti) self-dual YM equation which are everywhere non-singular and have
finite Euclidean YM actions on one hand while exhibiting meron-like large
distance behavior and carrying generally {\it fractional} topological
charge values on the other. Close inspection, however, reveals that the solutions
are more like instantons rather than merons in their generic natures. 

\end{abstract}

\pacs{PACS numbers: 11.15.-q, 04.40.-b, 04.60.-m \\
      Keywords: Yang-Mills instanton, Meron, Gravitational instanton}

\narrowtext

\begin{center}
{\rm \bf I. Introduction}
\end{center}

Certainly the discovery of the topologically degenerate vacuum structure of
non-abelian gauge theories was the starting point from which we began to
appreciate the fruitful but still mysterious non-perturbative regime of the
theories. And central to this non-perturbative aspects of non-abelian gauge
theories is the pseudoparticles, dubbed ``instantons'' [1]. In a naive
mathmatical sense, they are the classical solutions to Euclidean field equations
of non-abelian gauge theories and in a physical sense, they are the non-abelian
gauge field configurations interpolating between two homotopically distinct
but degenerate vacua. They thus can be thought of as saddle points which make
dominant contribution to the intervacua tunnelling amplitude in the path
integral formulation of quantum gauge theory. Of course the instanton physics
in pure non-abelian gauge theories such as Yang-Mills (YM) theory formulated in
flat Euclidean space has been studied thoroughly thus far. Its study in
non-trivial but physically meaningful gravitational fields, however, has been
extremely incomplete. Indeed, the strength of gravity well below the Planck
scale is negligibly small compared to those of elementary particle interactions
described by non-abelian gauge theories. Thus one might overlook the effects of
gravity on non-perturbative regime of non-abelian gauge theories such as the 
physics of instanton. Nevertheless, no matter how weak the relative strength of
the background gravity is, as long as the gravity carries non-trivial topology,
it may have profound effects on the structure of gauge theory instantons since
these instantons are topological objects linked to the topology-changing processes.
Therefore in the present work, we would like to explore how the topological
properties of the YM theory (or more precisely, of the YM instanton solution)
are dictated by the non-trivial topology  of the gravitational field with
which it interacts. Being an issue 
 of great physical interest and importance, quite a few serious study
 along this line have appeared in the literature but they were
 restricted to the background gravitational field with high degree
 of isometry such as the Euclideanized Schwarzschild
 geometry [2] or the Euclidean de Sitter space [3]. Even the works
 involving more general background spacetimes including gravitational
 instantons (GI) were mainly confined to the case of asymptotically-
 locally-Euclidean (ALE) spaces which is one particular such GI and
 employed rather indirect and mathmatically-oriented solution generating 
 methods such as the ADHM construction [14]. Recently, we [4] have 
 proposed a ``simply physical'' and hence perhaps the most direct
 algorithm for generating the YM instanton solutions in all species
 of known GI. Particularly, in [4] this new algorithm has been applied
 to the construction of solutions to (anti) self-dual YM equation in
 the background of Taub-NUT and Eguchi-Hanson metrics which are the 
 best-known such GI. In the present work, we would like to complete
 our discussion on this issue by providing a detailed presentation of
 our algorithm and applying it to practically all the GI known. The 
 careful physical interpretation of the solutions obtained eventually 
 to determine their nature will also be given in this work. 
 The essence of this method lies in writing the (anti)
 self-dual YM equation by employing truly relevant ans\H{a}tz for the
 YM gauge connection and then directly solving it. 
 To demonstrate how simple in method and powerful in applicability 
 it is, we then apply this algorithm to the case of (anti) self-dual YM
 equations in almost all of known GI and find the YM instanton solutions
 in their backgrounds. In particular, 
 the actual YM instanton solution in the background of Taub-NUT  
 (which is asymptotically-locally-flat (ALF) rather than ALE), Fubini-Study
 (on $CP^2$), and de Sitter (on $S^4$) metrics are
 constructed for the first time in this work.  Interestingly, the solutions to 
 (anti) self-dual YM equation turn out to be the rather exotic type of 
 instanton configurations which are everywhere non-singular having {\it finite}
 YM action but sharing some features with meron solutions [11] such as 
 their typical structure and generally {\it fractional} topological  charge values
 carried by them. Namely, the YM instanton solution that we shall discuss in
 the background of GI in this work exhibit characteristics which are mixture
 of those of typical instanton and typical meron. Thus at this point, it seems
 relevant to briefly review the essential nature of meron solution. For
 detailed description of meron, we refer the reader to some earlier works [10,11].
 First, recall that the standard BPST [1] SU(2) YM instanton solution in flat space 
takes the form $A^{a}_{\mu} = 2\eta^{a}_{\mu\nu}[x^{\nu}/(r^2+\lambda^2)]$
with $\eta^{a}_{\mu\nu}$ and $\lambda$ being the 'tHooft tensor [5] and the size of 
the instanton respectively while the meron solution which is another non-trivial
solution to the second order YM field equation found long ago by De Alfaro, Fubini, 
and Furlan [10] takes the form  $A^{a}_{\mu} = \eta^{a}_{\mu\nu}(x^{\nu}/r^2)$. 
Since the pure (vacuum) gauge having vanishing field strength is given by 
$A^{a}_{\mu} = 2\eta^{a}_{\mu\nu}(x^{\nu}/r^2)$, the standard instanton solution 
interpolates between the trivial vacuum $A^{a}_{\mu}=0$ at $r=0$ and another 
vacuum represented by this pure gauge above at $r\rightarrow \infty$ and the 
meron solution can be thought of as a ``half a vacuum gauge''. Unlike the instanton
solution, however, the meron solution only solves the second order YM field 
equation and fails to solve the first order (anti) self-dual equation. As is
apparent from their structures given above, the meron is an unstable solution
in that it is singular at its center $r=0$ and at $r=\infty $ while the ordinary
instanton solution exhibits no singular behavior. As was pointed out originally
by De Alfaro et al. [10], in contrast to instantons whose topological charge
density is a smooth function of $x$, the topological charge density of merons
vanishes everywhere except at its center, i.e., the singular point, such that
its volume integral is half unit of topological charge $1/2$. And curiously
enough, half-integer topological charge seems to be closely related to the
confinement in the Schwinger model [11]. It is also amusing to note that a
``time slice'' through the origin, i.e., $x_{0}=0$ of the meron configuration
yields a $SU(2)$ Wu-Yang monopole [11]. Lastly, the Euclidean meron action 
diverges logarithmically and perhaps needs some regularization whereas the 
standard YM instanton has finite action. \\
 We now recall some generic features of gravitational instantons. 
 In the loose sense, GI may be
 defined as a positive-definite metrics $g_{\mu\nu}$ on a complete
 and non-singular manifold satisfying the Euclidean Einstein
 equations and hence constituting the stationary points of the
 gravity action in Euclidean path integral for quantum gravity.
 But in the stricter sense [5,6], they are the metric solutions to the
 Euclidean Einstein equations having (anti) self-dual Riemann
 tensor
\begin{eqnarray}
\tilde{R}_{abcd} = {1\over 2}\epsilon_{ab}^{~~ef} R_{efcd} = \pm
R_{abcd}
\end{eqnarray}
(say, with indices written in non-coordinate orthonormal basis)
and include only two families of solutions in a rigorous sense ;
the Taub-NUT metric [7] and the Eguchi-Hanson instanton [8]. 
In the loose sense, however, there are several solutions to
Euclidean Einstein equations that can fall into the category of GI. 

\begin{center}
{\rm \bf II. New algorithm for solutions to (anti) self-dual YM equation}
\end{center}

We now begin with the action governing our
system, i.e., the Einstein-Yang-Mills (EYM) theory given by
\begin{eqnarray}
I_{EYM} = \int_{M} d^4x\sqrt{g}\left[{-1\over 16\pi}R + {1\over
4g^2_{c}}F^{a}_{\mu\nu}F^{a\mu\nu}\right] - \int_{\partial M}
d^3x\sqrt{h}{1\over 8\pi}K
\end{eqnarray}
where $F^{a}_{\mu\nu}$ is the field strength of the YM gauge field
$A^{a}_{\mu}$ with $a=1,2,3$ being the SU(2) group index and
$g_{c}$ being the gauge coupling constant. The Gibbons-Hawking
term on the boundary $\partial M$ of the manifold $M$ is also
added and $h$ is the metric induced on $\partial M$ and $K$ is the
trace of the second fundamental form on $\partial M$. Then by
extremizing this action with respect to the metric $g_{\mu\nu}$
and the YM gauge field $A^{a}_{\mu}$, one gets the following
classical field equations respectively
\begin{eqnarray}
&&R_{\mu\nu} - {1\over 2}g_{\mu\nu}R  =
8\pi T_{\mu\nu}, \nonumber \\
&&T_{\mu\nu} = {1\over
g^2_{c}} \left[F^{a}_{\mu\alpha}F_{\nu}^{a\alpha} - {1\over 4}
g_{\mu\nu}(F^{a}_{\alpha\beta}F^{a\alpha\beta})\right], \\
&&D_{\mu}\left[\sqrt{g}F^{a\mu\nu}\right] = 0, 
~~~D_{\mu}\left[\sqrt{g}\tilde{F}^{a\mu\nu}\right] = 0 \nonumber
\end{eqnarray}
where we added Bianchi identity in the last line and
$F^{a}_{\mu\nu} =
\partial_{\mu}A^{a}_{\nu}-\partial_{\nu}A^{a}_{\mu}+\epsilon^{abc}
A^{b}_{\mu}A^{c}_{\nu}$, $D^{ac}_{\mu} = \partial_{\mu}\delta^{ac}
+\epsilon^{abc}A^{b}_{\mu}$ and $A_{\mu}=A^{a}_{\mu}(-iT^{a})$,
$F_{\mu\nu}=F^{a}_{\mu\nu}(-iT^{a})$ with $T^{a}=\tau^{a}/2$
($a=1,2,3$) being the SU(2) generators and finally
$\tilde{F}_{\mu\nu} = {1\over 2}
\epsilon_{\mu\nu}^{~~\alpha\beta}F_{\alpha\beta}$ is the (Hodge) dual
of the field strength tensor. We now seek solutions ($g_{\mu\nu}$,
$A^{a}_{\mu}$) of the coupled EYM equations given above in
Euclidean signature obeying the (anti) self-dual equation in the
YM sector
\begin{eqnarray}
F^{\mu\nu} = g^{\mu\lambda}g^{\nu\sigma}F_{\lambda\sigma} = \pm
{1\over 2} \epsilon_{c}^{\mu\nu\alpha\beta}F_{\alpha\beta}
\end{eqnarray}
where
$\epsilon_{c}^{\mu\nu\alpha\beta}=\epsilon^{\mu\nu\alpha\beta}/\sqrt{g}$
is the curved spacetime version of totally antisymmetric tensor.
As was noted in [2,3], in Euclidean signature, the YM
energy-momentum tensor vanishes identically for YM fields
satisfying this (anti) self-duality condition. This point is of
central importance and can be illustrated briefly as follows.
Under the Hodge dual transformation, $F^{a}_{\mu\nu} \rightarrow
\tilde{F}^{a}_{\mu\nu}$, the YM energy-momentum tensor
$T_{\mu\nu}$ given in eq.(3) above is invariant normally in
Lorentzian signature. In Euclidean signature, however, its sign
flips, i.e., $\tilde{T}_{\mu\nu} = - T_{\mu\nu}$. As a result, for
YM fields satisfying the (anti) self-dual equation in Euclidean
signature such as the instanton solution, $F^{a}_{\mu\nu} = \pm
\tilde{F}^{a}_{\mu\nu}$, it follows that $T_{\mu\nu} =
-\tilde{T}_{\mu\nu} = -T_{\mu\nu}$, namely the YM energy-momentum
tensor vanishes identically, $T_{\mu\nu}=0$. This, then, indicates
that the YM field now does not disturb the geometry while the
geometry still does have effects on the YM field. Consequently the
geometry, which is left intact by the YM field, effectively serves
as a ``background'' spacetime which can be chosen somewhat at our
will (as long as it satisfies the vacuum Einstein equation
$R_{\mu\nu}=0$) and here in this work, we take it to be the
gravitational instanton. Loosely speaking, all the typical GI, including
Taub-NUT metric and Eguchi-Hanson solution, possess the same
topology $R\times S^3$ and similar metric structures. Of course in a
stricter sense, their exact topologies can be distinguished, say, by different
Euler numbers and Hirzebruch signatures [5,6]. Particularly,
in terms of the concise basis 1-forms, the metrics of these GI can
be written as [5,6]
\begin{eqnarray}
ds^2 &=& c^2_{r}dr^2 +
c^2_{1}\left(\sigma^2_{1}+\sigma^2_{2}\right) +
c^2_{3}\sigma^2_{3} \nonumber \\ &=& c^2_{r}dr^2 +
\sum_{a=1}^{3}c^2_{a}\left(\sigma^{a}\right)^2 = e^{A}\otimes
e^{A}
\end{eqnarray}
where $c_{r}=c_{r}(r)$, $c_{a}=c_{a}(r)$, $c_{1}=c_{2}\neq c_{3}$
and the orthonormal basis 1-form $e^{A}$ is given by
\begin{eqnarray}
e^{A} = \left\{e^{0}=c_{r}dr, ~~e^{a}=c_{a}\sigma^{a}\right\}
\end{eqnarray}
and $\left\{\sigma^{a}\right\}$ ($a=1,2,3$) are the left-invariant
1-forms satisfying the SU(2) Maurer-Cartan structure equation
\begin{eqnarray}
d\sigma^{a} = -{1\over 2}\epsilon^{abc}\sigma^{b}\wedge
\sigma^{c}.
\end{eqnarray}
They form a basis on the $S^{3}$ section of the geometry and hence
can be represented in terms of 3-Euler angles $0\leq \theta
\leq\pi$, $0\leq \phi \leq 2\pi$, and $0\leq \psi \leq 4\pi$
parametrizing $S^3$ as
\begin{eqnarray}
\sigma^1 &=& -\sin\psi d\theta + \cos\psi \sin\theta d\phi, \nonumber \\
\sigma^2 &=&  \cos\psi d\theta + \sin\psi \sin\theta d\phi, \\
\sigma^3 &=& -d\psi - \cos\theta d\phi. \nonumber
\end{eqnarray}
Now in order to construct exact YM instanton solutions in the
background of these GI, we now choose the relevant ans\H{a}tz for
the YM gauge potential and the SU(2) gauge fixing. And in doing
so, our general guideline is that the YM gauge field ans\H{a}tz
should be endowed with the symmetry inherited from that of the
background geometry, the GI. Thus we first ask what kind of
isometry these GI possess. As noted above, all the typical GI 
possess the topology of
$R\times S^3$. The geometrical structure of the $S^3$ section,
however, is not that of perfectly ``round'' $S^3$ but rather, that
of ``squashed'' $S^3$. In order to get a closer picture of this
squashed $S^3$, we notice that the $r=$constant slices of these GI
can be viewed as U(1) fibre bundles over $S^2\sim CP^1$ with the
line element
\begin{eqnarray}
d\Omega^2_{3} = c^2_{1}\left(\sigma^2_{1}+\sigma^2_{2}\right) +
c^2_{3}\sigma^2_{3} = c^2_{1}d\Omega^2_{2} +
c^2_{3}\left(d\psi + B\right)^2
\end{eqnarray}
where $d\Omega^2_{2}=(d\theta^2 + \sin^2\theta d\phi^2)$ is the
metric on unit $S^2$, the base manifold whose volume form
$\Omega_{2}$ is given by $\Omega_{2} = dB$ as $B = \cos\theta
d\phi$ and $\psi$ then is the coordinate on the U(1)$\sim S^1$
fibre manifold. Now then the fact that $c_{1}=c_{2}\neq c_{3}$
indicates that the geometry of this fibre bundle manifold is not
that of round $S^3$ but that of squashed $S^3$ with the squashing
factor given by $(c_{3}/c_{1})$. And further, it is squashed along
the U(1) fibre direction. Thus this failure for the geometry to be
that of exactly round $S^3$ keeps us from writing down the
associated ans\H{a}tz for the YM gauge potential right away.
Apparently, if the geometry were that of round $S^3$, one would
write down the YM gauge field ans\H{a}tz as $A^{a}=f(r)\sigma^{a}$
[3] with $\{\sigma^{a}\}$ being the left-invariant 1-forms introduced
earlier. The rationale for this choice can be stated
briefly as follows. First, since the $r=$constant sections of the
background space have the geometry of round $S^3$ and hence
possess the SO(4)-isometry, one would look for the SO(4)-invariant
YM gauge connection ans\H{a}tz as well. Next, noticing that both
the $r=$constant sections of the frame manifold and the SU(2) YM
group manifold possess the geometry of round $S^3$, one may
naturally choose the left-invariant 1-forms $\{\sigma^{a}\}$ as
the ``common'' basis for both manifolds. Thus this YM gauge
connection ans\H{a}tz, $A^{a}=f(r)\sigma^{a}$ can be thought of as
a hedgehog-type ans\H{a}tz where the group-frame index mixing is
realized in a simple manner [3]. Then coming back to our present
interest, namely the GI given in eq.(5), in $r=$constant sections,
the SO(4)-isometry is partially broken down to that of SO(3) by
the squashedness along the U(1) fibre direction to a degree set by
the squashing factor $(c_{3}/c_{1})$. Thus now our task became
clearer and it is how to encode into the YM gauge connection
ans\H{a}tz this particular type of SO(4)-isometry breaking coming
from the squashed $S^3$. Interestingly, a clue to this puzzle can
be drawn from the work of Eguchi and Hanson [9] in which they
constructed abelian instanton solution in Euclidean Taub-NUT
metric (namely the abelian gauge field with (anti)self-dual field
strength with respect to this metric). To get right to the point,
the working ans\H{a}tz they employed for the abelian gauge field
to yield (anti)self-dual field strength is to align the abelian
gauge connection 1-form along the squashed direction, i.e., along
the U(1) fibre direction, $A = g(r)\sigma^3$. This choice looks
quite natural indeed. After all, realizing that embedding of a
gauge field in a geometry with high degree of isometry is itself
an isometry (more precisly isotropy)-breaking action, it would be
natural to put it along the direction in which part of the
isometry is already broken. Finally therefore, putting these two
pieces of observations carefully together, now we are in the
position to suggest the relevant ans\H{a}tz for the YM gauge
connection 1-form in these GI and it is
\begin{eqnarray}
A^{a} = f(r)\sigma^{a} + g(r)\delta^{a3}\sigma^{3}
\end{eqnarray}
which obviously would need no more explanatory comments except
that in this choice of the ans\H{a}tz, it is implicitly understood
that the gauge fixing $A_{r}=0$ is taken. From this point on, the
construction of the YM instanton solutions by solving the
(anti)self-dual equation given in eq.(4) is straightforward. To
sketch briefly the computational algorithm, first we obtain the YM
field strength 2-form (in orthonormal basis) via exterior calculus
(since the YM gauge connection ans\H{a}tz is given in
left-invariant 1-forms) as $F^{a}=(F^{1}, F^{2}, F^{3})$ where
\begin{eqnarray}
F^{1} &=& {f'\over c_{r}c_{1}}(e^0\wedge e^1) + {{f[(f-1)+g]}\over
c_{2}c_{3}}(e^{2}\wedge e^{3}), \nonumber \\ 
F^{2} &=& {f'\over c_{r}c_{2}}(e^0\wedge e^2) + {{f[(f-1)+g]}\over
c_{3}c_{1}}(e^{3}\wedge e^{1}),  \\ 
F^{3} &=& {(f'+g')\over c_{r}c_{3}}(e^0\wedge e^3) + {{[f(f-1)-g]}\over
c_{1}c_{2}}(e^{1}\wedge e^{2}) \nonumber
\end{eqnarray}
from which we can read off the (anti)self-dual equation to be
\begin{eqnarray}
\pm {f'\over c_{r}c_{1}} = {{f[(f-1)+g]}\over c_{2}c_{3}},
~~~\pm {(f'+g')\over c_{r}c_{3}} = {{[f(f-1)-g]}\over c_{1}c_{2}} 
\end{eqnarray}
where the prime denotes the derivative with respect to $r$. After 
some manipulation, these (anti) self-dual equation can be cast to
a more practical form
\begin{eqnarray}
(\ln f)'' &+& \left[\left({c_{3}\over c_{r}}\right)'\left(
{c_{r}\over c_{3}}\right)\pm \left({c_{r}c_{3}\over
c_{1}c_{2}}\right)\right](\ln f)' =\left({c_{r}\over
c_{1}}\right)^{2}[f^2 - 1], \\ h &=& \left[1 \pm \left({c_{3}\over
c_{r}}\right)(\ln f)'\right]
\end{eqnarray}
where $h(r) = f(r)+g(r)$ and ``$+$'' for self-dual and ``$-$'' for
anti-self-dual equation and we have only a set of two equations as
$c_{1}=c_{2}$. Now the remaining computational algorithm is, for
each GI corresponding to particular choice of  $e^{A} =
\{e^{0}=c_{r}dr, ~~e^{a}=c_{a}\sigma^{a}\}$, first eqs.(13) and
(14), if admit solutions, give $f(r)$ and $g(r)$ respectively and
from which, next the YM instanton solutions in eq.(10) and their
(anti) self-dual field strength in eq.(11) can be obtained. At
this point, it is interesting to realize that actually there are 
other avenues to constructing the YM instanton solutions of
different species from that given in eq.(10) in these GI. To state
once again, in $r=constant$ sections of GI, since the
SO(4)-isometry is partially broken by the squashedness of $S^3$
along the U(1) fibre direction set by $\sigma^{3}$ in eq.(9), this
particular direction set by $\sigma_{3}$ can be thought of as a
kind of that of {\it principal axis}. Note also that exactly to   
the same degree this U(1) fibre direction set by $\sigma^{3}$
stands out, the other two directions set by $\sigma^{1}$ and
$\sigma^{2}$ respectively, may be regarded as being special. Thus
one might as well want to align the YM gauge connection solely   
along the direction set by $\sigma^{3}$ or along the direction set
by $\sigma^{1}$ or $\sigma^{2}$. And this can only be done when
one abandons the non-abelian structure in the YM gauge field and
writes its ans\H{a}tz in the form
\begin{eqnarray}
A^{a} = g(r)\delta^{a3}\sigma^{3} ~~~{\rm or} ~~~A^{a} =
g(r)\delta^{a1(2)}\sigma^{1(2)} \nonumber
\end{eqnarray}
respectively. Then YM instanton solutions of these species should
essentially be equivalent to the abelian instantons of the
Eguchi-Hanson-type mentioned earlier and as such they, if exist,
should clearly be totally different kinds of instanton solutions 
that cannot be related to the standard YM instantons given in
eq.(10) via any gauge transformation whatsoever. For this reason,
we shall call them ``abelianized'' YM instanton solutions and
attempt to construct them in this work as well. The field strength
and the (anti) self-dual equations associated with these
abelianized YM instantons are then given respectively by
\begin{eqnarray}
F^{a} &=& \left[{g'\over c_{r}c_{3}}(e^{0}\wedge e^{3}) - {g\over
c_{1}c_{2}}(e^{1}\wedge e^{2})\right]\delta^{a3}, \\ &\pm& (\ln
g)' = - \left({c_{r}c_{3}\over c_{1}c_{2}}\right) ~~~{\rm for}
~~~A^{a} = g(r)\delta^{a3}\sigma^{3} \nonumber
\end{eqnarray}
and
\begin{eqnarray}
F^{a} &=& \left[{g'\over c_{r}c_{1}}(e^{0}\wedge e^{1}) - {g\over 
c_{2}c_{3}}(e^{2}\wedge e^{3})\right]\delta^{a1}, \\ &\pm& (\ln
g)' = - \left({c_{r}\over c_{3}}\right) ~~~{\rm for} ~~~A^{a} =
g(r)\delta^{a1}\sigma^{1} \nonumber
\end{eqnarray}
and similarly for $A^{a} = g(r)\delta^{a2}\sigma^{2}$. And in the
above equations, ``$+$'' for self-dual and ``$-$'' for
anti-self-dual equation. We now present both the ``standard'' and
``abelianized'' YM instanton solutions of the forms given in
eqs.(10),(15), and (16) for each of the GI. 

\begin{center}
{\rm \bf III. Application of the algorithm to various GI backgrounds}
\end{center}

In this section, in order to exhibit how simple in method and how powerful
in applicability this new algorithm of ours really is, we shall apply the 
algorithm to the cases of Taub-NUT (TN), Eguchi-Hanson (EH), Fubini-Study (FS),
Taub-bolt (TB), and de Sitter GI backgrounds and find the solutions to (anti)
self-dual YM equations in these GI. \\   

{\rm \bf (1) YM instanton in Taub-NUT (TN) metric background} \\ 
The TN GI solution written in the metric form given in eq.(5) 
amounts to
\begin{eqnarray}
c_{r}={1\over 2}\left[{r+m\over r-m}\right]^{1/2},
~~~c_{1}=c_{2}={1\over 2}\left[r^2-m^2\right]^{1/2},
~~~c_{3}=m\left[{r-m\over r+m}\right]^{1/2} \nonumber
\end{eqnarray}
and it is a solution to Euclidean vacuum Einstein equation
$R_{\mu\nu}=0$ for $r\geq m$ with self-dual Riemann tensor. The
apparent singularity at $r=m$ can be removed by a coordinate
redefinition and is a `nut' (in terminology of Gibbons and Hawking
[6]) at which the isometry generated by the Killing vector
$(\partial/\partial \psi)$ has a zero-dimensional fixed point set.
The boundary of TN metric at $r\rightarrow \infty $ is $S^3$.
And this TN instanton is an asymptotically-locally-flat (ALF) metric. \\
(i) Standard YM instanton solution \\
It turns out that only the anti-self-dual equation
$F^{a}=-\tilde{F}^{a}$ admits a non-trivial solution and it is
$A^{a}=(A^1, A^2, A^3)$ where
\begin{eqnarray}
A^1 = \pm 2{(r-m)^{1/2}\over (r+m)^{3/2}}e^1,  ~~~A^2 = \pm
2{(r-m)^{1/2}\over (r+m)^{3/2}}e^2, ~~~A^3 = {(r+3m)\over m}
{(r-m)^{1/2}\over (r+m)^{3/2}}e^3
\end{eqnarray}
and $F^{a}=(F^1, F^2, F^3)$ where
\begin{eqnarray}
F^1 &=& \pm {8m\over (r+m)^3}\left(e^0\wedge e^1 - e^2\wedge
e^3\right), ~~~F^2 = \pm {8m\over (r+m)^3}\left(e^0\wedge e^2 -
e^3\wedge e^1\right), \nonumber \\ F^3 &=&  {16m\over
(r+m)^3}\left(e^0\wedge e^3 - e^1\wedge e^2\right).
\end{eqnarray}
It is interesting to note that this YM field strength and the
Ricci tensor of the background TN GI are proportional as
$|F^{a}|=2|R^{0}_{a}|$ except for opposite self-duality, i.e.,
\begin{eqnarray}
R^0_{1}=-R^2_3 &=&  {4m\over (r+m)^3}\left(e^0\wedge e^1 +
e^2\wedge e^3\right), ~~~R^0_{2}=-R^3_1 = {4m\over
(r+m)^3}\left(e^0\wedge e^2 + e^3\wedge e^1\right), \nonumber \\
R^0_{3}=-R^1_2 &=& -{8m\over (r+m)^3}\left(e^0\wedge e^3 +
e^1\wedge e^2\right).
\end{eqnarray}
(ii) Abelianized YM instanton along the direction set by $\sigma^3$ \\
Both the self-dual and anti-self-dual equations admit non-trivial
solutions and they are, in orthonormal basis,
\begin{eqnarray}
A^{a} &=& k\left({r+m\over r-m}\right)\delta^{a3}\sigma^{3} =
{k\over m}\left({r+m\over r-m}\right)^{3/2}\delta^{a3}e^{3}, \\
F^{a} &=& -{4k\over (r-m)^2}\left[(e^0\wedge e^3) + (e^1\wedge e^2)\right]
\delta^{a3} \nonumber
\end{eqnarray}
for the solution to self-dual equation and
\begin{eqnarray}
A^{a} &=& k\left({r-m\over r+m}\right)\delta^{a3}\sigma^{3} =
{k\over m}\left({r-m\over r+m}\right)^{1/2}\delta^{a3}e^{3}, \\  
F^{a} &=& {4k\over (r+m)^2}\left[(e^0\wedge e^3) - (e^1\wedge e^2)\right]
\delta^{a3} \nonumber
\end{eqnarray}
for the solution to anti-self-dual equation. In these solutions, $k$ is an
arbitrary constant. \\
(iii) Abelianized YM instanton along the direction set by $\sigma^1$ \\
Again, both the self-dual and anti-self-dual equations admit non-trivial
solutions and they are
\begin{eqnarray}
A^{a} &=& {k\over (r-m)}e^{-r/2m}\delta^{a1}\sigma^{1} =
{2k\over (r+m)^{1/2}(r-m)^{3/2}}e^{-r/2m}\delta^{a1}e^{1}, \\  
F^{a} &=& -{2k\over m}{e^{-r/2m}\over (r-m)^2}\left[(e^0\wedge e^1) +
(e^2\wedge e^3)\right]\delta^{a1} \nonumber
\end{eqnarray}
for the solution to self-dual equation and
\begin{eqnarray}
A^{a} &=& k(r-m)e^{r/2m}\delta^{a1}\sigma^{1} =
2k\left({r-m\over r+m}\right)^{1/2}e^{r/2m}\delta^{a1}e^{1}, \\
F^{a} &=& {2k\over m}e^{r/2m}\left[(e^0\wedge e^1) - (e^2\wedge e^3)\right]
\delta^{a1} \nonumber
\end{eqnarray}
for the solution to anti-self-dual equation. This solution, however, is {\it not}
physical and hence should be dropped as it blows up as $r\rightarrow \infty$. \\
{\rm \bf (2) YM instanton in Eguchi-Hanson (EH) metric background} \\ 
The EH GI solution amounts to
\begin{eqnarray}
c_{r}=\left[1 - \left({a\over r}\right)^4\right]^{-1/2},
~~~c_{1}=c_{2}={1\over 2}r, ~~~c_{3}={1\over 2}r\left[1 - \left({a\over
r}\right)^4 \right]^{1/2} \nonumber
\end{eqnarray}
and again it is a solution to Euclidean vacuum Einstein equation
$R_{\mu\nu}=0$ for $r\geq a$ with self-dual Riemann tensor. $r=a$
is just a coordinate singularity that can be removed by a
coordinate redefinition provided that now $\psi$ is identified
with period $2\pi$ rather than $4\pi$ and is a `bolt' (in
terminology of Gibbons and Hawking [6]) where the action of the
Killing field $(\partial/\partial \psi)$ has a two-dimensional
fixed point set. Note that for an ordinary $S^3$, the range for the
Euler angle $\psi$ would be $0\leq \psi \leq 4\pi$. Thus demanding 
$0\leq \psi \leq 2\pi$ instead to remove the bolt singularity at
$r=a$ amounts to identifying points antipodal with respect to the origin
and this, in turn, implies that the boundary of EH at $r\rightarrow \infty $
is the real projective space $RP^3 = S^3/Z_{2}$. Besides, this EH instanton is an
asymptotically-locally-Euclidean (ALE) metric. \\ 
(i) Standard YM instanton solution \\
In this time, only the self-dual equation $F^{a}=+\tilde{F}^{a}$ admits a non-trivial
solution and it is $A^{a}=(A^1, A^2, A^3)$ where
\begin{eqnarray}
A^1 = \pm {2\over r}\left[1 - \left({a\over r}\right)^4\right]^{1/2}e^1,
~~~A^2 = \pm {2\over r}\left[1 - \left({a\over r}\right)^4\right]^{1/2}e^2,
~~~A^3 = {2\over r}{\left[1 + \left({a\over r}\right)^4\right]\over \left[1 -
\left({a\over r}\right)^4\right]^{1/2}} e^3
\end{eqnarray}
and $F^{a}=(F^1, F^2, F^3)$ where
\begin{eqnarray}
F^1 &=& \pm {4\over r^2}\left({a\over r}\right)^4\left(e^0\wedge e^1 +
e^2\wedge e^3\right), ~~~F^2 = \pm {4\over r^2}\left({a\over
r}\right)^4\left(e^0\wedge e^2 + e^3\wedge e^1\right), \nonumber \\ F^3
&=& - {8\over r^2}\left({a\over r}\right)^4\left(e^0\wedge e^3 + e^1\wedge
e^2\right).
\end{eqnarray}
Again it is interesting to realize that this YM field strength and
the Ricci tensor of the background EH GI are proportional as
$|F^{a}|=2|R^{0}_{a}|$, i.e.,
\begin{eqnarray}
R^0_{1}=-R^2_3 &=& {2\over r^2}\left({a\over r}\right)^4\left(- e^0\wedge e^1
+ e^2\wedge e^3\right), ~~~R^0_{2}=-R^3_1 = {2\over r^2}\left({a\over
r}\right)^4\left(- e^0\wedge e^2 + e^3\wedge e^1\right), \nonumber \\
R^0_{3}=-R^1_2 &=& - {4\over r^2}\left({a\over r}\right)^4\left(- e^0\wedge
e^3 + e^1\wedge e^2\right).
\end{eqnarray}
It is also interesting to note that this YM instanton solution particularly
in EH background (which is ALE) obtained by directly solving the self-dual
equation can also be ``constructed'' by simply identifying 
$A^{a}=\pm 2\omega^{0}_{a}$ (where $\omega^{0}_{a}=(\epsilon_{abc}/2)\omega^{bc}$
are the spin connection of EH metric) and hence $F^{a}=\pm 2R^{0}_{a}$ as was
noticed by [13] but in the string theory context with different motivation.
This construction of solution via a simple identification of gauge field
connection with the spin connection, however, works only in ALE backgrounds
such as EH metric and generally fails as is manifest in the previous TN
background case (which is ALF, not ALE) in which $A^{a}\neq \pm 2\omega^{0}_{a}$
but still $F^{a}=\pm 2R^{0}_{a}$. Thus the method
presented here by first writing (by employing a relevant ans\H{a}tz for the YM gauge
connection given in eq.(10)) and directly solving the (anti) self-dual equation
looks to be the algorithm for generating the solution with general applicability
to all species of GI in a secure and straightforward manner. In this regard, the
method for generating YM instanton solutions to (anti) self-dual equation in
all known GI backgrounds proposed here in this work can be contrasted to 
earlier works in the literature [15] discussing the construction of YM instantons 
mainly in the background of ALE GI via indirect methods such as that of ADHM [14]. \\
(ii) Abelianized YM instanton along the direction set by $\sigma^3$ \\
Both the self-dual and anti-self-dual equations admit non-trivial
solutions and they are
\begin{eqnarray}
A^{a} &=& {k\over r^2}\delta^{a3}\sigma^{3} =
{2k\over r^3}\left[1 - \left({a\over r}\right)^4 \right]^{-1/2}\delta^{a3}e^{3}, \\
F^{a} &=& -{4k\over r^4}\left[(e^0\wedge e^3) + (e^1\wedge e^2)\right]
\delta^{a3} \nonumber
\end{eqnarray}  
for the solution to self-dual equation and
\begin{eqnarray}
A^{a} &=& kr^2 \delta^{a3}\sigma^{3} =
2kr\left[1 - \left({a\over r}\right)^4 \right]^{-1/2}\delta^{a3}e^{3}, \\
F^{a} &=& 4k \left[(e^0\wedge e^3) - (e^1\wedge e^2)\right]
\delta^{a3} \nonumber
\end{eqnarray}
for the solution to anti-self-dual equation. In these solutions, $k$ is an
arbitrary constant. Note, however, that this solution to the anti-self-dual equation
is {\it unphysical} and thus should be dropped as it fails to represent a localized
soliton configuration.\\
(iii) Abelianized YM instanton along the direction set by $\sigma^1$ \\
Again, both the self-dual and anti-self-dual equations admit non-trivial
solutions and they are
\begin{eqnarray}
A^{a} &=& {k\over \sqrt{r^4-a^4}}\delta^{a1}\sigma^{1} =
{2k\over r \sqrt{r^4-a^4}}\delta^{a1}e^{1}, \\   
F^{a} &=& -{4k\over (r^4-a^4)}\left[(e^0\wedge e^1) +
(e^2\wedge e^3)\right]\delta^{a1} \nonumber
\end{eqnarray}
for the solution to self-dual equation and
\begin{eqnarray}
A^{a} &=& k\sqrt{r^4-a^4}\delta^{a1}\sigma^{1} =
{2k\over r}\sqrt{r^4-a^4}\delta^{a1}e^{1}, \\
F^{a} &=& 4k\left[(e^0\wedge e^1) - (e^2\wedge e^3)\right]
\delta^{a1} \nonumber
\end{eqnarray}
for the solution to anti-self-dual equation. Again, this solution is {\it not}
physical and hence should be discarded as it fails to represent a localized soliton
configuration. \\
{\rm \bf (3) YM instanton in Fubini-Study (FS) metric on $CP^2$ background} \\
Lastly, the FS (on complex projective plane $CP^2$) gravitational instanton
solution corresponds to   
\begin{eqnarray}
c_{r}=\left[1 + {1\over 6}\Lambda r^2 \right]^{-1},
~~~c_{1}=c_{2}={r\over 2}\left[1 + {1\over 6}\Lambda r^2 \right]^{-1/2}, 
~~~c_{3}={r\over 2}\left[1 + {1\over 6}\Lambda r^2 \right]^{-1} \nonumber
\end{eqnarray}
where $\Lambda $ is the (positive) cosmological constant and it is a solution to
the Euclidean Einstein equation $R_{\mu \nu}=8\pi \Lambda g_{\mu \nu}$.
As such, this FS metric is a ``compact'' gravitational instanton (i.e., instanton
of finite volume) with no boundary and is everywhere regular up to the fact that 
a close inspection [5,6] reveals that at $r=0$, there is a removable nut singularity 
while at $r\rightarrow \infty $, we have a bolt singularity which is removable 
provided $0\leq \psi \leq 4\pi$. Besides, unlike the previous TN and
EH instantons which have self-dual Riemann tensors $R_{\mu\nu\alpha\beta}=
\tilde{R}_{\mu\nu\alpha\beta}$, this FS instanton possesses self-dual Weyl tensor
$C_{\mu\nu\alpha\beta}=\tilde{C}_{\mu\nu\alpha\beta}$. \\
(i) Standard YM instanton solution \\
Only the self-dual equation $F^{a}=+\tilde{F}^{a}$ admits a non-trivial solution
and the corresponding solution and the associated
self-dual field strength are given by
\begin{eqnarray}
A^1 = \pm {2\over r}e^1, ~~~A^2 = \pm {2\over r}e^2,
~~~A^3 = {2\over r}(1 + {1\over 12}\Lambda r^2)e^3
\end{eqnarray}
and $F^{a}=(F^1, F^2, F^3)$ where  
\begin{eqnarray}
F^1 &=& \pm {\Lambda \over 3}\left(e^0\wedge e^1 + e^2\wedge e^3\right), 
~~~F^2 = \pm {\Lambda \over 3}\left(e^0\wedge e^2 + e^3\wedge e^1\right), \nonumber \\ 
F^3 &=& - {\Lambda \over 3}\left(e^0\wedge e^3 + e^1\wedge e^2\right).
\end{eqnarray}
Again it is interesting to contrast this YM field strength with
the Ricci tensor of the background FS GI given by
\begin{eqnarray}
R^0_{1} &=& -R^2_3 = {\Lambda \over 6}\left(e^0\wedge e^1   
- e^2\wedge e^3\right), ~~~R^0_{2}=-R^3_1 = {\Lambda \over 6}
\left(e^0\wedge e^2 - e^3\wedge e^1\right), \nonumber \\   
R^0_{3} &=& {\Lambda \over 3}\left(2e^0\wedge e^3 + e^1\wedge e^2\right),
~~~R^1_{2} = {\Lambda \over 3}\left(e^0\wedge e^3 + 2e^1\wedge e^2\right)
\end{eqnarray}
which, unlike the TN and EH cases, {\it fails} to obey the relation $|F^{a}|=2|R^{0}_{a}|$ presumably
because the FS solution fails to have self-dual Riemann tensor.
Here it seems worthy of note that since the background FS metric is a compact instanton
and hence has a finite volume, one needs not worry about the possible divergence of the
field energy upon integration over the volume. Namely, this instanton solution is a
legitimate, physical solution. \\
(ii) Abelianized YM instanton along the direction set by $\sigma^3$ \\ 
Again, both the self-dual and anti-self-dual equations admit non-trivial
solutions and they are
\begin{eqnarray}
A^{a} &=& {6k\over \Lambda r^2}(1 + {1\over 6}\Lambda r^2)\delta^{a3}\sigma^{3} =
{12k\over \Lambda r^3}(1 + {1\over 6}\Lambda r^2)^2 \delta^{a3}e^{3}, \\
F^{a} &=& -{24k\over \Lambda r^4}(1 + {1\over 6}\Lambda r^2)^2
\left[(e^0\wedge e^3) + (e^1\wedge e^2)\right]\delta^{a3} \nonumber
\end{eqnarray}
for the solution to self-dual equation and
\begin{eqnarray}
A^{a} &=& {k\Lambda \over 6}r^2(1 + {1\over 6}\Lambda r^2)^{-1} \delta^{a3}\sigma^{3} =
{k\over 3}\Lambda r \delta^{a3}e^{3}, \\
F^{a} &=& {2k\Lambda \over 3} \left[(e^0\wedge e^3) - (e^1\wedge e^2)\right]
\delta^{a3} \nonumber
\end{eqnarray}
for the solution to anti-self-dual equation. In these solutions, $k$ is again an
arbitrary constant. \\
(iii) Abelianized YM instanton along the direction set by $\sigma^1$ \\
\begin{eqnarray}
A^{a} &=& {k\over r^2}\delta^{a1}\sigma^{1} =
{2k\over r^3}(1 + {1\over 6}\Lambda r^2)^{1/2}\delta^{a1}e^{1}, \\
F^{a} &=& -{4k\over r^4}(1 + {1\over 6}\Lambda r^2)^{3/2}\left[(e^0\wedge e^1) +
(e^2\wedge e^3)\right]\delta^{a1} \nonumber
\end{eqnarray}
for the solution to self-dual equation and
\begin{eqnarray}
A^{a} &=& kr^2 \delta^{a1}\sigma^{1} =
2kr(1 + {1\over 6}\Lambda r^2)^{1/2}\delta^{a1}e^{1}, \\
F^{a} &=& 4k(1 + {1\over 6}\Lambda r^2)^{3/2}\left[(e^0\wedge e^1) - (e^2\wedge e^3)\right] 
\delta^{a1}. \nonumber
\end{eqnarray}
for the solution to anti-self-dual equation.  \\
And this completes the presentation of all non-trivial YM instanton solutions in three 
families of gravitational instantons. We discussed earlier in the introduction the
classification of gravitational instantons [5,6]. And the three families of gravitational
instantons, TN, EH, and FS metrics fall into the class of instanton solutions in the
stricter sense as they have (anti) self-dual Riemann or Weyl tensor. In this classification, 
all the other gravitational instantons discovered thus far can be thought of as being
instanton solutions in the loose sense as they all fail to satisfy (anti) self-dual condition
for Riemann or Weyl tensor although still are the solutions to the Euclidean Einstein equation
with or without the cosmological constant. Therefore for the sake of completeness of our 
study, here we also provide explicit YM instanton solutions in the background of other
species of gravitational instantons in the loose sense. And particularly, we consider the
Taub-bolt metric [10] and the de Sitter metric on $S^4$ [5,6]. \\
{\rm \bf (4) YM instanton in Taub-bolt (TB) metric background} \\
This TB GI solutin written in the metric form given in eq.(5)
corresponds to
\begin{eqnarray}
c_{r}=\left[{2(r^2-N^2)\over 2r^2-5Nr+2N^2}\right]^{1/2},
~~~c_{1}=c_{2}=\left[r^2-N^2\right]^{1/2},
~~~c_{3}=2N\left[{2r^2-5Nr+2N^2 \over 2(r^2-N^2)}\right]^{1/2} \nonumber
\end{eqnarray}
and it is a solution to Euclidean vacuum Einstein equation
$R_{\mu\nu}=0$ for $r\geq 2N$. Again, in terminology of Gibbons and Hawking [5,6],
$r=2N$ is a `bolt' singularity that can be removed by a coordinate redefinition. As stated,
although neither its Riemann nor Weyl tensor is (anti) self-dual, it is, like the TN-metric,
another asymptotically locally-flat (ALF) instanton. \\
(i) Standard YM instanton solution \\
Unlike the ones belonging to the class of instanton solutions in the stricter sense, i.e.,
TN, EH, and FS metrics, neither self-dual nor anti-self-dual equation $F^{a}=\pm \tilde{F}^{a}$
in this TB-metric background admits any non-trivial solution. \\
(ii) Abelianized YM instanton along the direction set by $\sigma^3$ \\
Both the self-dual and anti-self-dual equations admit non-trivial
solutions and they are,
\begin{eqnarray}
A^{a} &=& k\left({r+N\over r-N}\right)\delta^{a3}\sigma^{3} =
{k\over 2N}\left({r+N\over r-N}\right)\left[{2(r^2-N^2)\over 2r^2-5Nr+2N^2}\right]^{1/2}
\delta^{a3}e^{3}, \nonumber \\  
F^{a} &=& -{k\over (r-N)^2}\left[(e^0\wedge e^3) + (e^1\wedge e^2)\right]
\delta^{a3} 
\end{eqnarray}
for the solution to self-dual equation and
\begin{eqnarray}
A^{a} &=& k\left({r-N\over r+N}\right)\delta^{a3}\sigma^{3} =
{k\over 2N}\left({r-N\over r+N}\right)\left[{2(r^2-N^2)\over 2r^2-5Nr+2N^2}\right]^{1/2}
\delta^{a3}e^{3}, \nonumber \\
F^{a} &=& {k\over (r+N)^2}\left[(e^0\wedge e^3) - (e^1\wedge e^2)\right]
\delta^{a3} 
\end{eqnarray}  
for the solution to anti-self-dual equation and where $k$ is an
arbitrary constant. \\
(iii) Abelianized YM instanton along the direction set by $\sigma^1$ \\
Again, both the self-dual and anti-self-dual equations admit non-trivial
solutions and they are
\begin{eqnarray}
A^{a} &=& {k\over (2r-N)^{1/4}(r-2N)}e^{-r/2N}\delta^{a1}\sigma^{1} =
{k\over (2r-N)^{1/4}(r-2N)(r^2-N^2)^{1/2}}e^{-r/2N}\delta^{a1}e^{1}, \nonumber \\
F^{a} &=& -{k\over \sqrt{2}N}{1\over (2r-N)^{3/4}(r-2N)^{3/2}}e^{-r/2N}
\left[(e^0\wedge e^1) + (e^2\wedge e^3)\right]\delta^{a1} 
\end{eqnarray}
for the solution to self-dual equation and
\begin{eqnarray}
A^{a} &=& k(2r-N)^{1/4}(r-2N)e^{r/2N}\delta^{a1}\sigma^{1} =
k{(2r-N)^{1/4}(r-2N)\over (r^2-N^2)^{1/2}}e^{r/2N}\delta^{a1}e^{1}, \nonumber \\
F^{a} &=& {k\over \sqrt{2}N}(2r-N)^{-1/4}(r-2N)^{1/2}e^{r/2N}
\left[(e^0\wedge e^1) - (e^2\wedge e^3)\right]\delta^{a1}
\end{eqnarray}
for the solution to anti-self-dual equation. Note, however, that this last 
solution is {\it unphysical} and hence should be discarded as it fails to
represent a localized soliton configuration. \\
{\rm \bf (5) YM instanton in the de Sitter metric on $S^4$ background} \\
This de Sitter (on $S^4$) gravitational instanton solution corresponds to
\begin{eqnarray}
c_{r}=\left[1+\left({r\over 2a}\right)^2\right]^{-1},
~~~c_{1}=c_{2}=c_{3}={r\over 2}\left[1+\left({r\over 2a}\right)^2\right]^{-1}
\end{eqnarray}
where $a$ is the radius of $S^4$ and it is a solution to Euclidean Einstein equation
$R_{\mu\nu}=8\pi \Lambda g_{\mu\nu}$. Thus the radius $a$ of $S^4$ is related to the 
inverse of $\sqrt{\Lambda}$ as $a=\sqrt{3/8\pi \Lambda}$. Like the FS metric we studied 
earlier, this de Sitter metric on $S^4$ is another {\it compact} gravitational instanton 
having no boundary and hence is 
everywhere regular. As is well-known, de Sitter space is a space of constant curvature
and hence is conformally-flat. Thus this de Sitter metric on $S^4$ has vanishing
Weyl tensor, $C_{\mu\nu\alpha\beta}=0$. This point is already evident from the fact that
$~c_{1}=c_{2}=c_{3}$ which indicates that the $r=$constant slices of this de Sitter
metric on $S^4$ geometry are {\it round} $S^3$'s with isometry group $SO(4)$. 
Thus the relevant ans\H{a}tz for YM gauge connection is simply
\begin{eqnarray}
A^{a} = f(r)\sigma^{a} \nonumber
\end{eqnarray}
for reasons stated earlier and the associated field strength and the (anti) self-dual
equation read
\begin{eqnarray}
F^{a} &=& {f'\over c_{r}c_{a}}(e^{0}\wedge e^{a}) + {1\over 2}\epsilon^{abc}
{f(f-1)\over c_{b}c_{c}}(e^{b}\wedge e^{c}), \nonumber \\
&&\pm {f'\over f(f-1)} = \left({c_{r}\over c_{1}}\right)
\end{eqnarray}
where ``$+$'' for self-dual and ``$-$'' for anti-self-dual equations. Obviously. this is
the special case when $g(r)=0$ in the more general case in eq.(10) we have been discussing.
Then the standard YM instanton solutions (the physical one) can be constructed in a quite
straightforward manner and they are
\begin{eqnarray}
A^{a} &=& \left[1+\left({r\over 2a}\right)^2\right]^{-1}\sigma^{a} = {2\over r}e^{a}, \\
F^{a} &=& -{1\over a^2}[(e^{0}\wedge e^{a}) + {1\over 2}\epsilon^{abc}(e^{b}\wedge e^{c})]
\nonumber 
\end{eqnarray}
for the solution to self-dual equation and
\begin{eqnarray}
A^{a} &=& \left[1+\left({2a\over r}\right)^2\right]^{-1}\sigma^{a} = {r\over 2a^2}e^{a}, \\
F^{a} &=& {(4a)^2\over r^4}[(e^{0}\wedge e^{a}) - {1\over 2}\epsilon^{abc}(e^{b}\wedge e^{c})]
\nonumber
\end{eqnarray}
for the solution to anti-self-dual equation. Note that these solutions in de Sitter (on $S^4$)
instanton background are legitimate instanton solutions, namely one needs not worry about
the seemingly divergent field energy upon integration over all space since the background 
de Sitter metric is a ``compact'' instanton with finite proper volume. 

\begin{center}  
{\rm \bf IV. Analysis of the nature of solutions to (anti) self-dual YM equation}
\end{center}

We now would like to examine the nature of the solutions to (anti) self-dual YM equation
in the background of various GI discussed in the previous section.  Among other things,
an interesting lesson we learned from this study is that, although expected
to some extent, the chances for the existence of standard YM instanton solutions
(to (anti) self-dual equations) get smaller as the degree of isometry owned by
each gravitational instanton gets lower from, say, the de Sitter GI to the ones
with self-dual Riemann or Weyl tensor and then next to the ones without.
Next, concerning the discovered structure of the $SU(2)$ YM instanton solutions
supported by these typical GI, there appears to be an interesting point worthy
of note. First, recall
that the relevant ans\H{a}tz for the YM gauge connection is of the
form $A^{a}=f(r)\sigma^{a}$ in the highly symmetric de Sitter instanton background
with topology of $R\times ({\rm round})S^3$ and of the
form $A^{a}=f(r)\sigma^{a} + g(r)\delta^{a3}\sigma^3$ in the less
symmetric GI backgrounds with topology of $R\times ({\rm
squashed})S^3$. Here, however, the physical interpretation of the nature of YM gauge
potential solutions $A^{a}$ is rather unclear when they are expressed in terms of the
left-invariant 1-forms $\{\sigma^{a}\}$ or the orthonormal basis $e^{A}$ in eq.(6).
Thus in order to get a better insight into the
physical meaning of the structure of these YM connection
ans\H{a}tz, we now try to re-express the left-invariant 1-forms
$\{\sigma^{a}\}$ forming a basis on $S^3$ in terms of more
familiar Cartesian coordinate basis. And this can be achieved
by first relating the polar coordinates $(r, ~\theta, ~\phi, ~\psi)$ to
Cartesian $(t,x,y,z)$ coordinates (note, here, that $t$ is not
the usual ``time'' but just another spacelike coordinate) given by
[5]
\begin{eqnarray}
x+iy = r\cos {\theta\over 2}\exp{[{i\over 2}(\psi+\phi)]}, ~~~z+it =
r\sin {\theta\over 2}\exp{[{i\over 2}(\psi-\phi)]},
\end{eqnarray}
where $x^2+y^2+z^2+t^2=r^2$ which is the equation for $S^3$ with radius $r$.
From this coordinate transformation law, one now can relate the 
non-coordinate basis to the Cartesian coordinate basis as
\begin{eqnarray}
\pmatrix{dr \cr
         r\sigma_{x} \cr
         r\sigma_{y} \cr 
         r\sigma_{z} \cr} = {1\over r}
\pmatrix{x & y & z & t \cr
         -t & -z & y & x \cr
         z & -t & -x & y \cr
         -y & x & -t & z \cr}
\pmatrix{dx \cr
         dy \cr
         dz \cr
         dt \cr}
\end{eqnarray}
where $\{\sigma_{x}=-\sigma^{1}/2, ~\sigma_{y}=-\sigma^{2}/2,
 ~\sigma_{z}=-\sigma^{3}/2\}$.
Still, however, the meaning of YM gauge connection ans\H{a}tz rewritten
in terms of the Cartesian coordinate basis $dx^{\mu}=(dt, ~dx, ~dy, ~dz)$
as above does not look so apparent. Thus we next introduce
the so-called {\it `tHooft tensor} [1,11] defined by
\begin{eqnarray}
\eta^{a\mu\nu}=-\eta^{a\nu\mu}=(\epsilon^{0a\mu\nu}+{1\over 2}\epsilon^{abc}
\epsilon^{bc\mu\nu}).
\end{eqnarray} 
Then the left-invariant 1-forms can be cast to a more concise form
$\sigma^{a}=2\eta^{a}_{\mu\nu}(x^{\nu}/r^2)dx^{\mu}$. Therefore,
the YM instanton solution, in Cartesian coordinate basis, can be
written as
\begin{eqnarray}
A^{a} = A^{a}_{\mu}dx^{\mu} =
2\left[f(r)+g(r)\delta^{a3}\right]\eta^{a}_{\mu\nu}{x^{\nu}\over
r^2}dx^{\mu}
\end{eqnarray}
in the background of TN, EH, FS, and TB GI with topology of $R\times ({\rm
squashed})S^3$. Now in order to appreciate the meaning of this structure,
we go back to the flat space situation. As is well-known, 
the standard BPST [1] SU(2) YM instanton solution in flat space takes the
form $A^{a}_{\mu} = 2\eta^{a}_{\mu\nu}[x^{\nu}/(r^2+\lambda^2)]$
with $\lambda$ being the size of the instanton. Recall, however,
that separately from this BPST instanton solution, there is
another non-trivial solution to the YM field equation of
the form $A^{a}_{\mu} = \eta^{a}_{\mu\nu}(x^{\nu}/r^2)$ found long
ago by De Alfaro, Fubini, and Furlan [10]. (Note that the pure gauge
is given by $A^{a}_{\mu} = 2\eta^{a}_{\mu\nu}(x^{\nu}/r^2)$. Thus
the ordinary instanton solution interpolates between the trivial
vacuum $A^{a}_{\mu}=0$ at $r=0$ and another vacuum represented by the 
pure gauge above at $r\rightarrow \infty$ and the meron solution can
be thought of as a ``half a vacuum gauge''.)  This second solution is
called ``meron'' [11] as it carries a half unit of topological charge 
and is known to play a certain role
concerning the quark confinement [11]. It, however, exhibits
singularity at its center $r=0$ and hence has a diverging action and falls 
like $1/r$ as $r\rightarrow \infty$. Thus we are led to the conclusion 
that the YM instanton solution in typical GI backgrounds possess the
structure of (curved space version of) meron at large $r$. As is well-known,
in flat spacetime meron does not solve the 1st order (anti) self-dual 
equation although it does the second order YM field equation.
Thus in this sense, this result seems remarkable since it implies 
that in the GI backgrounds, the (anti) self-dual YM equation admits 
solutions which exhibit the configuration of meron solution at large $r$
in contrast to the flat spacetime case. And we only
conjecture that when passing from the flat ($R^4$) to GI ($R\times
S^3$) geometry, the closure of the topology of part of the
manifold appears to turn the structure of the instanton solution
from that of standard BPST into that of meron. The concrete form of
the YM instanton solutions in each of these GI backgrounds written in
terms of Cartesian coordinate basis as in eq.(49) will be given below after
we comment on one more thing. \\
Finally, we turn to investigation of other physical quantities such as 
the topological charge of each of these solutions and the 
estimate of the instanton contributions to the intervacua tunnelling 
amplitude which can serve as crucial indicators in determining the true 
physical natures of these solutions. It has been pointed out in the
literature that both in the background of Euclidean Schwarzschild
geometry [2] and in the Euclidean de Sitter space [3], the (anti)
instanton solutions have the Pontryagin index of $\nu[A]= \pm 1$
and hence give the contribution to the (saddle point approximation
to) intervacua tunnelling amplitude of $\exp{[-8\pi^2/g^2_{c}]}$,
which, interestingly, are the same as their flat space
counterparts even though these curved space YM instanton solutions
do not correspond to gauge transformations of any flat space
instanton solution [1]. This unexpected and hence rather curious
property, however, turns out not to persist in YM instantons  
in these GI backgrounds we studied here. In order to see this, 
we begin with the careful definition of the Pontryagin index or
second Chern class in the presence of the non-trivial background
geometry of GI. \\
Consider that we would like to find an index theorem for the
manifold ($M$) with boundary ($\partial M$). Namely, we now need 
an extended version of index theorem with boundary. To this question,
an appropriate answer has been provided by Atiyah, Patodi, and
Singer (APS) [12]. According to their extended version of index theorem,
the total index, say, of a given geometry and of a gauge field
receives contributions, in addition to that from the usual bulk 
term ($V(M)$), from a local boundary term ($S(\partial M)$) and
from a non-local boundary term ($\xi (\partial M)$). The bulk term
is the usual term appearing in the ordinary index theorem without
boundary and involves the integral over $M$ of terms 
quadratic in curvature tensor of the geometry and in field strength 
tensor of the gauge field. The local boundary term is given by the 
integral over $\partial M$ of the Chern-Simons forms for both the
geometry and the gauge field while the non-local boundary term
is given by a constant times the ``APS $\eta$-invariant'' [5] of
the boundary. And this last non-local boundary term becomes relevant
and meaningful when Dirac spinor field is present and interacts with
the geometry and the gauge field. Now specializing to the case at hand in
which we are interested in the evaluation of the instanton 
number or the second Chern class of the YM gauge field {\it alone},  
we only need to pick up the terms in the gauge sector in this APS
index theorem which reads [5]
\begin{eqnarray}
\nu[A] = Ch_{2}(F) = {-1\over 8\pi^2}[\int_{M=R\times S^3}tr(F\wedge F) -
\int_{\partial M=S^3}tr(\alpha \wedge F)|_{r=r_{0}}] 
\end{eqnarray}
where $\alpha \equiv (A-A')$ is the ``second fundamental form'' {\it at} the
boundary $r=r_{0}$ and by definition [5] $A'$ has only {\it tangential} 
components on the boundary $\partial M=S^3$. Recall, however, that our
choice of ans\H{a}tz for the YM gauge connection involves the gauge fixing
$A_{r}=0$ as we mentioned earlier. Namely, both $A$ and $A'$ possess only
tangential components (with respect to the $r=r_{0}$ boundary) at any 
$r=r_{0}$ and hence $\alpha \equiv (A-A') = 0$ identically there. As a result,
even in the presence of the boundaries, the terms in the YM gauge sector
in the APS index theorem remain the same as in the case of index theorem with
no boundary, namely, only the bulk term survives in eq.(50) above. Thus what remains
is just a straightforward computation of this bulk term and it becomes easier
when performed in terms of orthonormal basis 
$e^{A} = \left\{e^{0}=c_{r}dr, ~~e^{a}=c_{a}\sigma^{a}\right\}$, in which case,
\begin{eqnarray}
tr(F\wedge F) &=& {1\over 2}(F^{a}\wedge F^{a}) = {1\over 2}({1\over 4})
\epsilon_{ABCD}F^{a}_{AB}F^{a}_{CD} \sqrt{g}d^4x \nonumber \\
&=& (F^{1}_{01}F^{1}_{23}+F^{2}_{02}F^{2}_{31}+F^{3}_{03}F^{3}_{12})\sqrt{g}d^4x, \\
\int_{M=R\times S^3}d^4x\sqrt{g} &=& \int_{R}dr (c_{r}c_{1}c_{2}c_{3})\int^{4\pi}_{0}
d\psi \int^{2\pi}_{0}d\phi \int^{\pi}_{0}d\theta \sin \theta \nonumber \\
&=& 16\pi^{2} \int_{R} dr (c_{r}c_{1}c_{2}c_{3}) \nonumber
\end{eqnarray}
where we used $\sqrt{g}=|det e|=c_{r}c_{1}c_{2}c_{3}\sin \theta$. The period for the
$U(1)$ fibre coordinate $\psi$ for the EH metric, however, is $2\pi$ rather than 
$4\pi$ to remove the bolt singularity at $r=a$ as we mentioned earlier. This completes
the description of the method for computing the topological charge of each solution.
Our next job, then, is the estimate of the instanton contributions to the intervacua
tunnelling amplitudes. Generally, the saddle point approximation to the intervacua 
tunnelling amplitude is given by
\begin{eqnarray}
\Gamma_{GI} \sim \exp{[-I_{GI}(instanton)]}
\end{eqnarray}
where the subscript ``GI'' denotes corresponding quantities in
the GI backgrounds and $I_{GI}(instanton)$ represents the
Euclidean YM theory action evaluated at the YM instanton solution,
i.e.,
\begin{eqnarray}
I_{GI}(instanton) = \int_{R\times S^3}d^4x\sqrt{g}\left[{1\over
4g^2_{c}}F^{a}_{\mu\nu}F^{a\mu\nu}\right] = \left({8\pi^2\over
g^2_{c}}\right)|\nu[A]|
\end{eqnarray}
where we used $4tr(F\wedge F) = F^{a}_{\mu\nu}\tilde{F}^{a \mu\nu}\sqrt{g}d^4x$
and the (anti)self-duality relation $F^{a}=\pm \tilde{F}^{a}$. 
The calculation of the Pontryagin indices and hence
the Euclidean YM actions we just described is indeed quite straightforward. \\
In the following, as we promised, we now provide the expression for the YM 
instanton solutions in each of these GI backgrounds written in terms of Cartesian 
coordinate basis to study its structure one by one in detail and also
we demonstrate the explicit evaluation of the topological charge values
and the estimate of the contributions to the intervacua
tunnelling amplitude in order eventually to determine the physical nature 
of each solution. \\
{\rm \bf (1) YM instanton in Taub-NUT metric background} \\
In terms of the ans\H{a}tz functions $f(r)$ and $g(r)$ for the YM gauge connection 
in GI backgrounds given in eq.(10), the standard instanton solutions in TN metric
amount to 
\begin{eqnarray}
f(r) &=& \left({r-m \over r+m}\right), ~~~g(r) = \left({2m\over r+m}\right)
\left({r-m \over r+m}\right), \\
f(r) &=& -\left({r-m \over r+m}\right), ~~~g(r) = 2\left({r+2m\over r+m}\right)
\left({r-m \over r+m}\right) \nonumber
\end{eqnarray}
for self-dual and anti-self-dual YM equations respectively.
Therefore, when expressed in Cartesian coordinate basis as in eq.(49), 
the solutions take the forms
\begin{eqnarray}
A^{a}_{\mu} &=& 2\left({r-m \over r+m}\right)\left[1 + \left({2m\over r+m}\right)
\delta^{a3}\right]\eta^{a}_{\mu\nu}{x^{\nu}\over r^2}, \\
A^{a}_{\mu} &=& 2\left({r-m \over r+m}\right)\left[-1 + 2\left({r+2m\over r+m}\right)
\delta^{a3}\right]\eta^{a}_{\mu\nu}{x^{\nu}\over r^2} \nonumber
\end{eqnarray}
for self-dual and anti-self-dual case respectively. Some comments regarding the
features of these solutions are now in order. i) They appear to be singular at the
center $r=0$ but it should not be a problem as $r\geq m$ for the background TN metric
and hence the point $r=0$ is absent. ii) It is interesting to note that the solutions
become vacuum gauge $A^{a}_{\mu}=0$ at the boundary $r=m$ which has the topology of
$S^3$. iii) For $r\rightarrow \infty$, the solutions asymptote to another vacuum gauge 
$|A^{a}_{\mu}| = 2\eta^{a}_{\mu\nu}(x^{\nu}/r^2)$. \\
We now turn to the computation of the topological charge, i.e., the Pontryagin index
of these YM solution. The relevant quantities involved in this computation are the
ones in eq.(51) and they, for the case at hand, are
\begin{eqnarray}
(c_{r}c_{1}c_{2}c_{3}) &=& {m\over 8}(r^2-m^2), \\
F^{a}_{\mu\nu}\tilde{F}^{a \mu\nu} &=& 
4(F^{1}_{01}F^{1}_{23}+F^{2}_{02}F^{2}_{31}+F^{3}_{03}F^{3}_{12}) 
= -24{(8m)^2\over (r+m)^6}. \nonumber
\end{eqnarray}
Thus we have
\begin{eqnarray}
\nu[A] &=& \left({-1\over 32\pi^2}\right)16\pi^2 \int^{\infty}_{m}dr {m\over 8}(r^2-m^2)
\left[-24{(8m)^2\over (r+m)^6}\right] \nonumber \\
&=& 1.
\end{eqnarray}
Then next the Euclidean YM action evaluated at these instanton solutions and hence the
saddle point approximation to the intervacua tunnelling amplitude are given 
respectively by
\begin{eqnarray}
I_{GI}(instanton) &=&  \left({8\pi^2\over g^2_{c}}\right)|\nu[A]|
= {8\pi^2\over g^2_{c}}, \\
\Gamma_{GI} \sim  \exp{[-I_{GI}(instanton)]} &=& \exp{(-8\pi^2/g^2_{c})}. \nonumber
\end{eqnarray}
{\rm \bf (2) YM instanton in Eguchi-Hanson metric background} \\
The standard instanton solutions in EH metric amount to
\begin{eqnarray}
f(r) &=& \left[1-\left({a\over r}\right)^4\right]^{1/2}, 
~~~g(r) = \left[1+\left({a\over r}\right)^4\right] -
\left[1-\left({a\over r}\right)^4\right]^{1/2}, \\
f(r) &=& -\left[1-\left({a\over r}\right)^4\right]^{1/2}, 
~~~g(r) = \left[1+\left({a\over r}\right)^4\right] +
\left[1-\left({a\over r}\right)^4\right]^{1/2}  \nonumber  
\end{eqnarray}  
for self-dual and anti-self-dual YM equations respectively.
Thus in Cartesian coordinate basis, the solutions take the forms
\begin{eqnarray}
A^{a}_{\mu} &=& 2\left\{\left[1-\left({a\over r}\right)^4\right]^{1/2} +
\left(\left[1+\left({a\over r}\right)^4\right] -
\left[1-\left({a\over r}\right)^4\right]^{1/2}\right)\delta^{a3}\right\}
\eta^{a}_{\mu\nu}{x^{\nu}\over r^2}, \\
A^{a}_{\mu} &=& 2\left\{-\left[1-\left({a\over r}\right)^4\right]^{1/2} +        
\left(\left[1+\left({a\over r}\right)^4\right] +
\left[1-\left({a\over r}\right)^4\right]^{1/2}\right)\delta^{a3}\right\}
\eta^{a}_{\mu\nu}{x^{\nu}\over r^2} \nonumber
\end{eqnarray}
for self-dual and anti-self-dual cases respectively. Some comments regarding the
features of these solutions are now in order. i) Again, they appear to be singular at the
center $r=0$ but it should not be a problem as $r\geq a$ for the background EH metric 
and hence the point $r=0$ is absent. ii) The solutions become
$A^{a}_{\mu}=4\eta^{a}_{\mu\nu}\delta^{a3}(x^{\nu}/r^2)$ at the boundary $r=a$ which 
has the topology of $S^3/Z_{2}$. 
iii) For $r\rightarrow \infty$, the solutions asymptote to the vacuum gauge
$|A^{a}_{\mu}| = 2\eta^{a}_{\mu\nu}(x^{\nu}/r^2)$. \\
We turn now to the computation of the Pontryagin index of these YM solution. 
For the case at hand, the relevant quantities involved in this computation are
\begin{eqnarray}
(c_{r}c_{1}c_{2}c_{3}) = {1\over 8}r^3, 
~~~F^{a}_{\mu\nu}\tilde{F}^{a \mu\nu} = 24\left({4a^4\over r^6}\right)^2. 
\end{eqnarray}
Thus we have
\begin{eqnarray}
\nu[A] = \left({-1\over 32\pi^2}\right)8\pi^2 \int^{\infty}_{a}dr {1\over 8}r^3
\left[24\left({4a^4\over r^6}\right)^2\right] 
= -{3\over 2}
\end{eqnarray}
where we set the range for the $U(1)$ fibre coordinate as $0\leq \psi \leq 2\pi$
rather than $0\leq \psi \leq 4\pi$ for the reason stated earlier.
Note particularly that it is precisely this point that renders the Pontryagin index
of this solution {\it fractional} because otherwise, it would come out as $-3$ instead.
Then next the Euclidean YM action evaluated at these instanton solutions and hence the
saddle point approximation to the intervacua tunnelling amplitude are given
respectively by
\begin{eqnarray}
I_{GI}(instanton) &=&  \left({8\pi^2\over g^2_{c}}\right)|\nu[A]|
= {12\pi^2\over g^2_{c}}, \\
\Gamma_{GI} \sim  \exp{[-I_{GI}(instanton)]} &=& \exp{(-12\pi^2/g^2_{c})}. \nonumber
\end{eqnarray}
{\rm \bf (3) YM instanton in Fubini-Study metric on $CP^2$ background} \\
The standard instanton solutions in FS metric amount to
\begin{eqnarray}
f(r) &=& \left[1+{1\over 6}\Lambda r^2\right]^{-1/2},
~~~g(r) = {[1+\Lambda r^2/12]\over [1+\Lambda r^2/6]} -
{1\over [1+\Lambda r^2/6]^{1/2}}, \\
f(r) &=& -\left[1+{1\over 6}\Lambda r^2\right]^{-1/2},
~~~g(r) = {[1+\Lambda r^2/12]\over [1+\Lambda r^2/6]} +
{1\over [1+\Lambda r^2/6]^{1/2}} \nonumber
\end{eqnarray}
for self-dual and anti-self-dual YM equations respectively.
Then in terms of Cartesian coordinate basis, the solutions take the forms
\begin{eqnarray}
A^{a}_{\mu} &=& {2\over [1+\Lambda r^2/6]^{1/2}}
\left\{1 + \left[{(1+\Lambda r^2/12)\over (1+\Lambda r^2/6)^{1/2}} 
- 1\right]\delta^{a3}\right\} \eta^{a}_{\mu\nu}{x^{\nu}\over r^2}, \\
A^{a}_{\mu} &=& {2\over [1+\Lambda r^2/6]^{1/2}}
\left\{-1 + \left[{(1+\Lambda r^2/12)\over (1+\Lambda r^2/6)^{1/2}}    
+ 1\right]\delta^{a3}\right\} \eta^{a}_{\mu\nu}{x^{\nu}\over r^2} \nonumber 
\end{eqnarray}
for self-dual and anti-self-dual cases respectively. Now, note that :
i)  The solution to the self-dual YM equation $A^{a}_{\mu} = 2\eta^{a}_{\mu\nu}(x^{\nu}/r^2)$
looks singular at the center $r=0$ since $0\leq r <\infty $ for the background FS
metric and hence the point $r=0$ is present. But this is just a pure gauge representing
a vacuum and thus should not be a trouble. Next, the solution to the anti-self-dual YM 
equation, $|A^{1,2}_{\mu}| = -2\eta^{1,2}_{\mu\nu}(x^{\nu}/r^2)$ and
$|A^{3}_{\mu}| = 2\eta^{3}_{\mu\nu}(x^{\nu}/r^2)$ is again a pure gauge having a vanishing
field strength. ii) For $r\rightarrow \infty$, the solutions asymptote to 
$|A^{1,2}_{\mu}| = 2\sqrt{6/\Lambda} \eta^{1,2}_{\mu\nu}(x^{\nu}/r^3)\rightarrow 0$ and
$|A^{3}_{\mu}| = \eta^{3}_{\mu\nu}(x^{\nu}/r^2)$ which is a component of flat space
meron solution. \\
Turning  now to the calculation of the Pontryagin index of these YM solutions,
again the relevant quantities involved in this computation are
\begin{eqnarray}
(c_{r}c_{1}c_{2}c_{3}) = {r^3\over 8(1+{1\over 6}\Lambda r^2)^{3}}, 
~~~F^{a}_{\mu\nu}\tilde{F}^{a \mu\nu} = {4\over 3}\Lambda^2 . \nonumber
\end{eqnarray}
Thus 
\begin{eqnarray}
\nu[A] = \left({-1\over 32\pi^2}\right)16\pi^2 \int^{\infty}_{0}dr 
{r^3\over 8(1+{1\over 6}\Lambda r^2)^{3}} \left[{4\over 3}\Lambda^2\right] 
= -{3\over 4}.
\end{eqnarray}
Next the Euclidean YM action and the
saddle point approximation to the intervacua tunnelling amplitude are given
respectively by
\begin{eqnarray}
I_{GI}(instanton) &=&  \left({8\pi^2\over g^2_{c}}\right)|\nu[A]|
= {6\pi^2\over g^2_{c}}, \\
\Gamma_{GI} \sim  \exp{[-I_{GI}(instanton)]} &=& \exp{(-6\pi^2/g^2_{c})}. \nonumber
\end{eqnarray}
{\rm \bf (4) YM instanton in the de Sitter metric on $S^4$ background} \\
In terms of the ans\H{a}tz functions $f(r)$ for the YM gauge connection    
given earlier as $A^{a} = f(r)\sigma^{a}$,
the standard instanton solutions in de Sitter metric amount to
\begin{eqnarray}
f(r) = \left[1+\left({r\over 2a}\right)^2\right]^{-1}, 
~~~f(r) = \left[1+\left({2a\over r}\right)^2\right]^{-1}  
\end{eqnarray}
for self-dual and anti-self-dual YM equations respectively.
Then in terms of Cartesian coordinate basis, the solutions take the forms
\begin{eqnarray}
A^{a}_{\mu} = {2\over [1+(r/2a)^2]}\eta^{a}_{\mu\nu}{x^{\nu}\over r^2}, 
~~~A^{a}_{\mu} = {2\over [1+(2a/r)^2]}\eta^{a}_{\mu\nu}{x^{\nu}\over r^2}
\end{eqnarray}
for self-dual and anti-self-dual cases respectively. Now, note that :
i) The solution to the self-dual YM equation $A^{a}_{\mu} = 2\eta^{a}_{\mu\nu}(x^{\nu}/r^2)$
is again a pure gauge representing a vacuum and thus should not be a trouble although the
point $r=0$ is present in the background de Sitter space. The solution to the 
anti-self-dual YM equation also approaches the vacuum, i.e., 
$A^{a}_{\mu} \simeq (1/2a^2)\eta^{a}_{\mu\nu}x^{\nu}\rightarrow 0$.
ii) For $r\rightarrow \infty$, the solutions asymptote to
$A^{a}_{\mu} = 8a^2 \eta^{a}_{\mu\nu}(x^{\nu}/r^4) \sim 0$ for the self-dual case and
$A^{a}_{\mu} = 2\eta^{a}_{\mu\nu}(x^{\nu}/r^2)$ which is the pure gauge for the
anti-self-dual case. \\
Lastly, turning to the calculation of the Pontryagin index of these YM solutions,
we first obtain the relevant quantities involved in this computation which are
\begin{eqnarray}
(c_{r}c_{1}c_{2}c_{3}) &=& {r^3\over 8[1+(r/2a)^2]^4}, \\
F^{a}_{\mu\nu}\tilde{F}^{a \mu\nu} &=& {1\over 2}\epsilon_{ABCD}F^{a}_{AB}F^{a}_{CD}
= {12\over a^4}. \nonumber 
\end{eqnarray}
Thus
\begin{eqnarray}
\nu[A] = \left({-1\over 32\pi^2}\right)16\pi^2 \int^{\infty}_{0}dr
{r^3\over 8[1+(r/2a)^2]^4} \left[{12\over a^4}\right] = -1.
\end{eqnarray}
Then the Euclidean YM action and the
saddle point approximation to the intervacua tunnelling amplitude are given
respectively by
\begin{eqnarray}
I_{GI}(instanton) &=&  \left({8\pi^2\over g^2_{c}}\right)|\nu[A]|
= {8\pi^2\over g^2_{c}}, \\
\Gamma_{GI} \sim  \exp{[-I_{GI}(instanton)]} &=& \exp{(-8\pi^2/g^2_{c})}. \nonumber
\end{eqnarray}
Let us now discuss the behavior of these solutions as $r\rightarrow 0$ once again
to stress that they really do not exhibit singular 
behaviors there. For TN, EH, and TB instantons, the
ranges for radial coordinates are $m\leq r <\infty$, $a\leq r
<\infty$, and  $2N\leq r <\infty$ respectively.
Since the point $r=0$ is absent in these
manifolds, the solutions in these GI are everywhere
regular. For the rest of the ``compact'' gravitational instantons, i.e.,
FS on $CP^2$ and de Sitter on $S^4$, however, the radial coordinate
runs $0\leq r <\infty$. Thus the point $r=0$ indeed is present in these   
compact instantons. The solutions in FS and de Sitter backgrounds, however,
seem to have no trouble either as they are essentially vacuum gauges having
vanishing field strength there at $r=0$. At large $r$, on the other hand, all
the solutions appear to take the structure close to that of meron solution
in flat space. Another interesting point worthy of note is
that the solutions in TN and de Sitter backgrounds exhibit a generic
property of the instanton solution in that they do interpolate between a vacuum
at $r=m$ ($r=0$) and another vacuum at $r\rightarrow \infty $.
Namely, the solutions in these GI backgrounds appear to exhibit
features of both meron such as their large $r$ behavior 
and instanton such as interpolating configurations between two vacua in some cases. 
Next, we analyze the meaning of the topological charge values of
the solutions and their contributions to the intervacua tunnelling
amplitudes. Except for the solutions in the background of TN metric
and de Sitter metric, generally the solutions in other GI backgrounds
such as EH and FS carry {\it fractional} topological charges smaller
or greater than unity in magnitude. Here, however, the
solution in EH metric background carries the half-integer Pontryagin
index actually because the range for the $U(1)$ fibre coordinate is
$0\leq \psi \leq 2\pi$ and hence the boundary of EH space is $S^3/Z_2$.
For the FS (on $CP^2$) case, it is unclear what is the true origin 
for the fractional Pontryagin index. Therefore
the fact that solutions in GI backgrounds generally carry fractional topological
charges appears to be another manifestation for mixed instanton-meron nature
of the solutions. Thus to summarize, the solutions in
TN and de Sitter backgrounds particularly display features generic in the
standard instanton while in the case of those in EH and FS backgrounds, such
generic features of the instanton is somewhat obscured by meron-type natures.
There, however, is one obvious consensus. All the solutions in these GI
backgrounds are non-singular at their centers and have finite Euclidean
YM action. And this last point allows us to suspect that these solutions
are more like instantons in their generic nature although looks rather like
merons in their structures. 

\begin{center}
{\rm\bf V. Concluding remarks}
\end{center}

We now summarize the results and close with some comments.
As we stressed earlier in the introduction, when it comes to the topological
aspect, gravity may have marked effects even at the level of elementary
particle physics despite its negligibly small relative strength well below
the Planck scale. Although this intriguing possibility has been pointed out
long ago, surprisingly little attempt has been made toward the demonstration of
this phenomenon in relevant physical systems. Thus in the present work, we 
took a concrete step toward in this direction. Namely, we attempted to construct
in an explicit and precise manner the $SU(2)$ YM instanton solutions practically
in all known gravitational instanton backgrounds. And in doing so, the task of
solving coupled Einstein-Yang-Mills equations for the metric and YM gauge field
has been greatly simplified by the fact that in Euclidean signature, the YM field
does not disturb the geometry as its energy-momentum tensor vanishes identically
as long as one looks only for the YM instanton solutions having (anti) self-dual
field strength.  Among other things,
an interesting lesson we learned from this study is that, although expected
to some extent, the chances for the existence of standard YM instanton solutions
(to (anti) self-dual equations) get smaller as the degree of isometry owned by
each gravitational instanton gets lower from, say, the de Sitter GI to the ones
with self-dual Riemann or Weyl tensor and then next to the ones without.
As demonstrated, it is also intersting to note that
the solutions turn out to take the structure of merons at large $r$ and
generally carry fractional topological charge values. Nevertheless, it
seems more appropriate to conclude that the solutions still should be
identified with (curved space version of) instantons as they
are solutions to 1st order (anti) self-dual equation and are everywhere
regular having finite YM action. However, these curious mixed characteristics
of the solutions to (anti) self-dual YM equation in GI backgrounds appear
to invite us to take them more seriously and further explore
potentially interesting physics associated with them.

\begin{center}
{\rm\bf Acknowledgments}
\end{center}

This work was supported in part by the Brain Korea 21 project and by the basic
science promotion program from Korea Research Foundation. Yoon also wishes to acknowledge
financial support of Hanyang univ. made in the program year of 1999.

\vspace*{2cm}

\noindent

\begin{center}
{\rm\bf References}
\end{center}

\begin{description}

\item {[1]} A. A. Belavin, A. M. Polyakov, A. S. Schwarz, and Yu. S. Tyupkin,
            Phys. Lett. {\bf B59}, 85 (1975) ;
            G. `tHooft, Phys. Rev. Lett. {\bf 37}, 8 (1976).
\item {[2]} J. M. Charap and M. J. Duff, Phys. Lett. {\bf B69}, 445 (1977) ;
            {\it ibid} {\bf B71}, 219 (1977).
\item {[3]} H. Kim and S. K. Kim,
            Nuovo Cim. {\bf B114}, 207 (1999) and references therein.
\item {[4]} H. Kim and Y. Yoon, Phys. Lett. {\bf B495}, 169 (2000) ({\it hep-th/0002151}).
\item {[5]} T. Eguchi, P. B. Gilkey, and A. J. Hanson, Phys. Rep. {\bf 66}, 213 (1980).
\item {[6]} G. W. Gibbons and C. N. Pope, Commun. Math. Phys. {\bf 66}, 267 (1979) ;
            G. W. Gibbons and S. W. Hawking, {\it ibid}, {\bf 66}, 291 (1979).
\item {[7]} A. Taub, Ann. Math. {\bf 53}, 472 (1951) ;
            E. Newman, L. Tamburino, and T. Unti, J. Math. Phys. {\bf 4}, 915 (1963) ;
            S. W. Hawking, Phys. Lett. {\bf A60}, 81 (1977).
\item {[8]} T. Eguchi and A. J.Hanson, Phys. Lett. {\bf B74}, 249 (1978).
\item {[9]} T. Eguchi and A. J. Hanson, Ann Phys. {\bf 120}, 82 (1979).
\item {[10]} V. De Alfaro, S. Fubini, and G. Furlan, Phys. Lett. {\bf B65}, 163 (1976).
\item {[11]} C. G. Callan, R. Dashen, and D. J. Gross, Phys. Rev. {\bf D17}, 2717 (1978).
\item {[12]} M. F. Atiyah, V. K. Patodi, and I. M. Singer, Bull. London Math. Soc. {\bf 5},
             229 (1973) ; Proc. Camb. Philos. Soc. {\bf 77}, 43 (1975) ; {\it ibid.} {\bf 78},
             405 (1975) ; {\it ibid.} {\bf 79}, 71 (1976).
\item {[13]} M. Bianchi, F. Fucito, G. C. Rossi, and M. Martellini, Nucl. Phys. {\bf B440},
             129 (1995).
\item {[14]} M. Atiyah, V. Drinfeld, N. Hitchin, and Y. Manin, Phys. Lett. {\bf A65}, 185
             (1987) ; P. B. Kronheimer, and H. Nakajima, Math. Ann. {\bf 288}, 263 (1990).
\item {[15]} H. Boutaleb-Joutei, A. Chakrabarti, and A. Comtet, Phys. Rev. {\bf D20}, 1844
             (1979) ; {\it ibid.} {\bf D20}, 1898 (1979) ; {\it ibid.} {\bf D21}, 979 (1980) ; 
             {\it ibid.} {\bf D21}, 2280 (1980) ; A. Chakrabarti, Fortschr. Phys. {\bf 35}, 1
             (1987) ;  M. Bianchi, F. Fucito, G. C. Rossi, and M. Martellini, Phys. Lett.
             {\bf B359}, 49-61 (1995).

\end{description}

\end{document}